\documentclass{article} 
\usepackage{iclr2023_conference,times}

\usepackage{amsmath,amsfonts,bm}

\def\eqref#1{equation~\ref{#1}}

\def\1{\bm{1}}

\DeclareMathAlphabet{\mathsfit}{\encodingdefault}{\sfdefault}{m}{sl}
\SetMathAlphabet{\mathsfit}{bold}{\encodingdefault}{\sfdefault}{bx}{n}

\usepackage[utf8]{inputenc} 
\usepackage[T1]{fontenc}    
\usepackage{hyperref}       
\usepackage{url}            
\usepackage{booktabs}       
\usepackage{amsfonts}       
\usepackage{nicefrac}       
\usepackage{microtype}      
\usepackage{xcolor}         

\usepackage{multirow}
\usepackage{graphicx}
\usepackage{amsmath,amsfonts,amsthm}
\usepackage{mathtools}
\usepackage{algorithm}
\usepackage[noend]{algorithmic}
\usepackage{algorithmic}
\usepackage{adjustbox}
\usepackage{multicol}
\usepackage{multirow}
\usepackage{amsmath}
\usepackage{caption}
\usepackage{subcaption}
\usepackage{wrapfig}

\theoremstyle{plain}

\theoremstyle{definition}

\theoremstyle{remark}

\usepackage{bbding}
\usepackage{wrapfig}

\title{Secure Federated Learning against Model Poisoning Attacks via Client Filtering}
\iclrfinalcopy

\author{%
  Duygu Nur Yaldiz$\phantom{}^{*}$
  , Tuo Zhang$\phantom{}^{*}$, Salman Avestimehr\\
  Department of Electrical and Computer Engineering\\
  University of Southern California \\
  \texttt{yaldiz@usc.edu, tuozhang@usc.edu, avestime@usc.edu} \\
}

\footnotetext[1]{The first two authors contributed equally.}

\begin{document}

\maketitle

\begin{abstract}
Given the distributed nature, detecting and defending against the backdoor attack under federated learning (FL) systems is challenging. In this paper, 
we observe that the cosine similarity of the last layer's weight between the global model and each local update could be used effectively as an indicator of malicious model updates. Therefore, we propose \texttt{CosDefense}, a cosine-similarity-based attacker detection algorithm. Specifically, under \texttt{CosDefense}, the server calculates the cosine similarity score of the last layer's weight between the global model and each client update, labels malicious clients whose score is much higher than the average, and filters them out of the model aggregation in each round. Compared to existing defense schemes, \texttt{CosDefense} does not require any extra information besides the received model updates to operate and is compatible with client sampling. Experiment results on three real-world datasets demonstrate that \texttt{CosDefense} could provide robust performance under the state-of-the-art FL poisoning attack.
\end{abstract}

\section{Introduction}

The gist of Federated Learning (FL) is to train a model coordinated by a server while preserving the clients' data privacy~\cite{Zhang2021FederatedLF}. However, this substantial property introduces new challenges. Since the server does not have access to the client data due to privacy concerns, FL is vulnerable to data or model poisoning attacks, in which the attacker send corrupted updates and contaminates the global model. Given the distributed nature of FL, it is challenging to detect and correct these failures under the vanilla FL framework~\cite{fedavg, zhang2022fedaudio}.

Several solutions have been proposed to defend the server from model poisoning attacks to relax the security challenge for the FL framework. Server-side robust aggregation approaches aim to detect outliers by inspecting the client updates, and filtering the malicious updates before model aggregation such as \cite{Krum}. 
Besides completely filtering out before model aggregation, approaches proposed by \cite{RFFL, FLTrust, ByGARS, FoolsGold, Prakash2020SecureAF} diminish the aggregation coefficients of the clients that are likely to be malicious. 
However, existing approaches have some critical drawbacks or unfeasible assumptions as we summarize in Table \ref{comparison-table}.


In this work, we propose \texttt{CosDetect}, a cosine similarity based outlier detection algorithm, to tackle the fundamental issues of existing defense methods. We provide an intriguing finding that the weight inside the last layer of the local model update is more sensitive to the local data distribution than other layers. Based on this crucial observation, we propose that the last layer of local updates from the malicious clients should be outliers compared to the ones from the benign clients. By calculating the cosine similarity of the last layer between each collected model update and the last global model, it is possible to filter the poisoning updates before model aggregation.

\begin{table*}[h]
\caption{Comparison of \texttt{CosDefense} with existing backdoor defense FL methods.}
\label{comparison-table}
\begin{center}
\begin{scriptsize}
\begin{tabular}{l|cccc}
\toprule

& \begin{tabular}{c}\textbf{Compatibility with}\\\textbf{Client Sampling}\\\end{tabular} 
&\begin{tabular}{c}\textbf{Validation Data}\\\textbf{at the Server}\\\end{tabular} 
& \begin{tabular}{c}\textbf{Information of}\\\textbf{Attacker Number}\\\end{tabular} 
& \begin{tabular}{c}\textbf{Client Score}\\\textbf{Maintenance}\\\textbf{at the Server}\\\end{tabular} \\

\midrule 
Krum \cite{Krum}          & $\surd$  & $\times$  & $\surd$  & $\times$  \\
Multi-Krum \cite{Krum}    & $\surd$  & $\times$  & $\surd$  & $\times$  \\
Median \cite{trimmed-mean}& $\surd$  & $\times$  & $\times$ & $\times$  \\
RFFL \cite{RFFL}          & $\times$ & $\times$  & $\times$ & $\surd$  \\
FoolsGold \cite{FoolsGold}& $\surd$  & $\times$  & $\times$ & $\surd$  \\
FLTrust \cite{FLTrust}    & $\surd$  & $\surd$   & $\times$ & $\times$ \\
ByGars \cite{ByGARS}      & $\surd$  & $\surd$   & $\times$ & $\surd$  \\
SageFlow \cite{Sageflow}  & $\surd$  & $\surd$   & $\times$ & $\times$  \\
\textbf{CosDefense (Our Method)}   & \CheckmarkBold  & \XSolidBold  & \XSolidBold & \XSolidBold \\
\bottomrule
\end{tabular}
\end{scriptsize}
\end{center}
\vskip -0.1in
\end{table*}

As shown in Table~\ref{comparison-table}, the proposed cosine similarity based outlier detection scheme, though simple, has equipped \texttt{CosDetect} with multiform merits compared to prior strategies: (1) \texttt{CosDetect} does not require representative benign data at the server to distinguish the malicious updates. It is only built on accessing the global model parameters and the clients' updates. (2) \texttt{CosDetect} does not need the precise number of malicious clients per communication round in advance for robust performance. (3) \texttt{CosDetect} does not rely on the previous clients' records to cluster the malicious clients. Instead, \texttt{CosDetect} eliminates the outliers only based on the current round information, which makes our algorithm compatible with client selection and brings more flexibility for the implementation.

\section{Approach Overview}

\subsection{Layer-wise Cosine Similarity for Model Weights}

\begin{wrapfigure}{r}{0.35\textwidth}
  \begin{center}
	\includegraphics[width=0.35\textwidth]{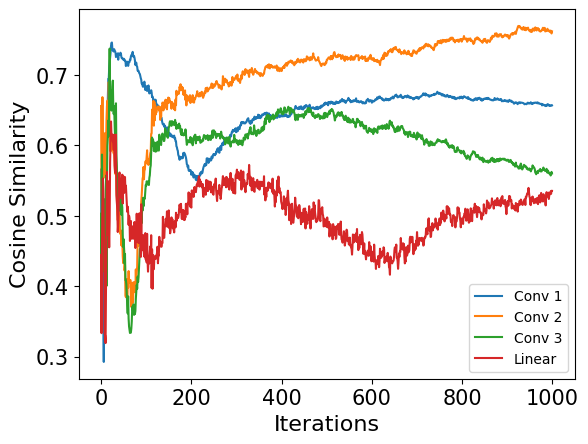}
    \caption{The average cosine similarity of model weight in various layers between one client and the other nine clients. All 10 models are trained independently for 1000 iterations without synchronization.}
    \label{fig:cos}
  \end{center}
\end{wrapfigure}

We first focus on a critical question: \textit{how would the attack be reflected on the server side during FL training?} Under the vanilla FL setting~\cite{fedavg}, the only information the server holds during the training are the collected model updates and the following aggregated model. A previous study~\cite{Zhao2020iDLGID} notes that label information for the training data can be computed analytically from the gradients of the last layer inside the machine learning model under the centralized setting. Following this direction, one intriguing finding from our empirical study is that compared to other layers, the last layer's weight is more sensitive to the input data distribution. We quantify the similarity of model weights by cosine similarity, which is the dot product of the two vectors divided by the product of individual norms as 
$ cos(\alpha) = \frac{<x, y>}{||x|| \cdot ||y||} $
where $\alpha$ is the angle between vectors $x$ and $y$. Figure~\ref{fig:cos} shows the average cosine similarity for each layer in the model across independent clients. In this experiment, ten clients train a four-layer CNN-based model independently without any model synchronization on the MNIST dataset, which has been non-iid partitioned (See Section~\ref{sec:exp set} for details). We observe that with the increment of the iteration, the input-side layers show higher similarity than the output-side layers, and the last layer has the lowest similarity score, because the local data distribution among all clients mainly varies on the label distribution. These observations provide critical insight that the local data label distribution could be efficiently reflected in the last layer's weight compared to the other layers.

\subsection{CosDefense: Filtering the Malicious Model Update via Cosine Similarity}

Based on the crucial observation we described in the previous section, we propose that the last layer of the local models from the attackers should be outliers compared to the ones from the benign models. Therefore, it is possible to filter the malicious model update on the server by calculating the cosine similarity between each collected model update and the last global model.
When calculating the cosine similarity between global model weights and local model update, we have the following:
\begin{align}
    cos(\alpha_i) &= \frac{<\theta_t, g_{t+1}^i>}{||\theta_t|| \cdot ||g_{t+1}^i||} 
    = \frac{<\theta_t, \theta_{t+1}^i - \theta_t>}{||\theta_t|| \cdot ||\theta_{t+1}^i - \theta_t||} 
    = \frac{<\theta_t, \theta_{t+1}^i> - <\theta_t, \theta_t>}{||\theta_t|| \cdot ||\theta_{t+1}^i - \theta_t||}
\end{align}
where $\alpha_i$ denotes the angle between global model weights ($\theta_t$) and local model update of client $i$ ($g_{t+1}^i$). If all the participated clients are benign nodes, as the communication round $t$ goes to infinity, the global model should be converged to an optimal point based on the received local model update. Therefore, the cosine similarity between global model weights and the local model update tends to be decreasing during the training, as the difference between $\theta_t$ and $\theta_{t+1}^i$ (local model for client $i$) goes smaller for each client. 

However, the cosine similarity scores from malicious clients do not follow this phenomenon. As their aim is to prevent convergence, their local optimization direction would be different compared to the benign client models or the global model. As a result, the difference between $\theta_t$ and $\theta_{t+1}^i$ from malicious nodes becomes more extensive than the benign nodes, making the cosine similarity score between the malicious client update and the global weights tend to be increasing during the training. 

\begin{wrapfigure}{r}{0.35\textwidth}
  \begin{center}
    \includegraphics[width=0.35\textwidth]{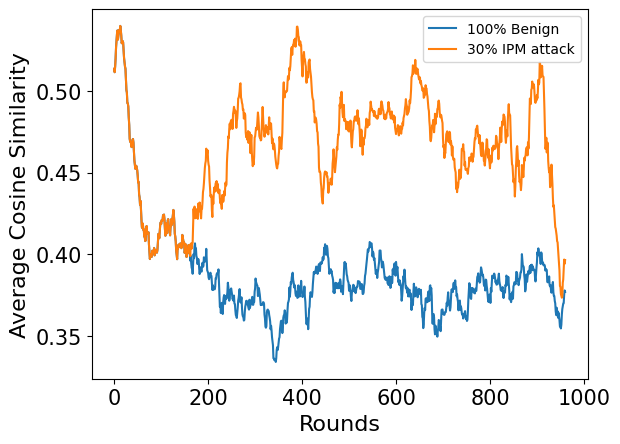}
    \caption{The trace of the average cosine similarity score of the last layer's weight between all received local updates and global model. }
    \label{fig:cos2}
  \end{center}
  \vskip -0.1in
\end{wrapfigure}

To demonstrate our thoughts, we trace the average cosine score of the last layer local updates and that of the global model under benign and poisoning attack cases ( Figure~\ref{fig:cos2}). Under both cases, the total client number is 100, and the sampling rate is 0.1 in each round. For the poisoning attack case, we launch the IPM attack~\cite{ipm} at round 200, with 30\% of the participants being malicious clients. For a better illustration, Figure~\ref{fig:cos2} is smoothed by the moving averaged filter with window size 40. The average cosine similarity increases sharply as the attack happens and is much higher than the scores under the all-benign case, which confirms our speculation.

We observed that cosine similarity between global model parameters and the benign local model updates tends to be smaller during the training, which indicates that the direction of benign updates is less than perpendicular to the global model, making it move toward a different direction in each round. However, the average cosine similarity scores increase after the attack, indicating that cosine similarity scores from malicious clients are much larger than the scores from benign clients. The large cosine similarity values represent that malicious client updates are aligned with the global model parameters and do not update the global model in different directions but keep it still. As a result, if there are enough malicious clients in the system, their updates considerably diminish the benefit of benign client updates, preventing the convergence of the global model.

Following these directions, we propose \texttt{CosDefense}, a server-side clustering-based defense method as shown in the Appendix \ref{algorithm}. In each round $t$, after the server receives all the updates from the sampled clients, it calculates the cosine similarity between the last layer's global weights and the last layer's local updates for each client. Then, the server clusters clients based on their cosine similarity scores either as malicious or benign. If the cosine similarity score of a client is much higher than others, then the server labels that client as malicious for round $t$ and exclude that client from aggregation. 
%
\texttt{CosDefense} only detects malicious clients before aggregation and excludes them, allowing the server to perform any aggregation method with benign client updates. Hence, \texttt{CosDefense} algorithm is compatible with any aggregation method.

\section{Evaluations}
\subsection{Experiment Setup} \label{sec:exp set}
We conduct the evaluations on three datasets and two models: MNIST~\cite{MNIST} and Fashion-MNIST~\cite{FMNIST} with a four-layer CNN-based model following the previous related work~\cite{Rl-attack}, and CIFAR-10~\cite{Cifar10} with ResNet-18 model~\cite{He2015DeepRL}. We generate the non-iid partition for all datasets following previous FL works~\cite{Rl-attack, Fang2019LocalMP} with default $q$ as 0.5, where the higher $q$ represents a higher non-iid degree. We compare our defense method with two representative robust aggregation rules: Krum~\cite{Krum} and Clipping-Median. We do not include the baselines such as FLTrust~\cite{FLTrust} or RFFL~\cite{RFFL} because they either need validation data on the server or do not compatible with the client sampling. Further details of the baseline implementation and parameter selection can be found in Appendix \ref{app:exp_details}

\subsection{Defense Performance with the State-of-the-Art Model Poisoning Attack} \label{sec:defense-perf}

To evaluate the defense performance, we evaluate the \texttt{CosDefense} on top of the state-of-the-art model poisoning attack methods, Inner Product Manipulation (IPM) \cite{ipm} attack. 
In the following experiments, we set the attacker control 30\% of the participants as malicious nodes, and the attack would launch at round 200. The server randomly samples 10 clients from 100 for local model training in each communication round.

\begin{figure*}[h!]
\centering
\begin{tabular}{ccc}
  \includegraphics[width=40mm]{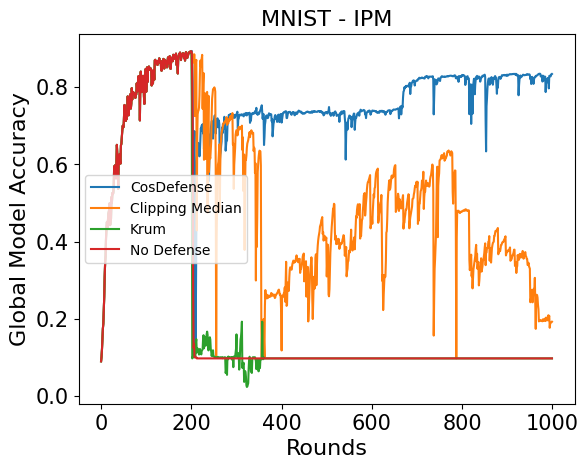} & \includegraphics[width=40mm]{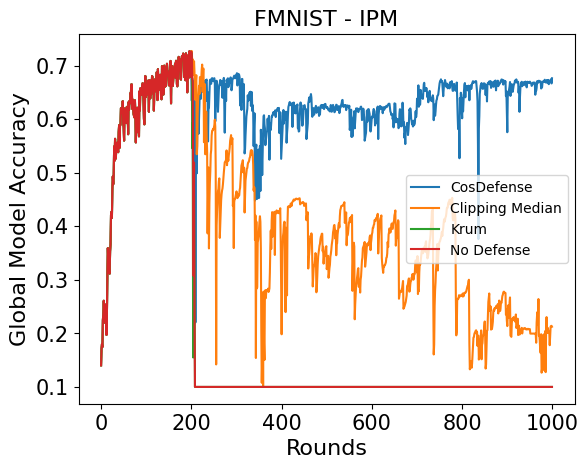}  &   \includegraphics[width=40mm]{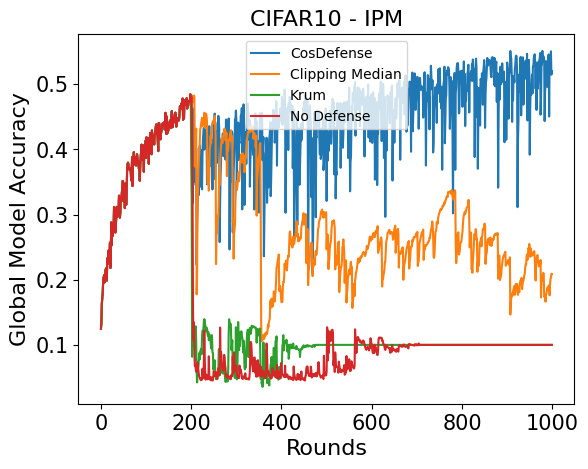}\\
\end{tabular}
\caption{Accuracy performance over 1000 rounds for Krum, Clipping-Median, \texttt{CosDefense}, and FedAvg on MNIST (left), FMNIST (middle), and CIFAR-10 (right) under IPM attack. }
\label{results-6}
\end{figure*}

\textbf{Evaluation Results:} Figure~\ref{results-6} shows the accuracy curve over 1000 communication rounds for Krum, Clipping-Median, and \texttt{CosDefense} for MNIST, Fashion-MNIST, and CIFAR-10. The detailed numerical results and ablation study related to the non-iid degree are shown in Appendix~\ref{app:result}.
We have three observations from the results. (1) As Krum and Clipping-Median use all layers of the updates, the results of \texttt{CosDefense} indicate that the similarity of the last layer between the collected model update and global model is more sensitive to attacks, which could effectively be used to filter out the malicious model updates. (2) After the IPM attack begins, all the defense strategies drop the accuracy significantly. However, the proposed \texttt{CosDefense} strategy has a much faster convergence speed and shorter recovery time to resume the learning compared to Krum and Clipping-Median. (3) For the most challenging dataset CIFAR-10, after the IPM attack happens, \texttt{CosDefense} not only recovers the best accuracy before round 200, but also has an increasing trend on the curve, while other defense baselines even could not reach the best accuracy before the attack.

\begin{wrapfigure}{r}{0.3\textwidth}
  \begin{center}
    \includegraphics[width=0.3\textwidth]{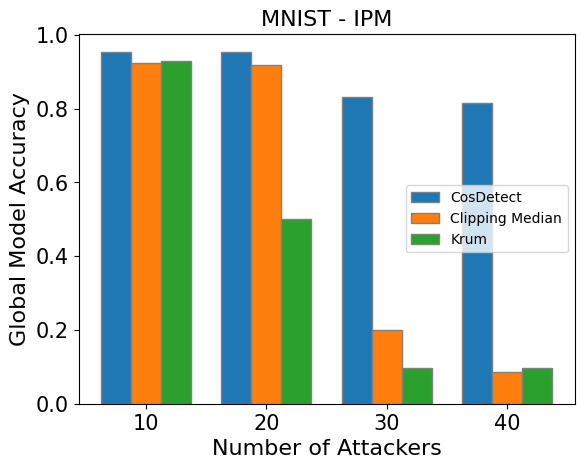}
    \caption{Accuracy performance on MNIST dataset under different numbers of attackers perform IPM attacks.}
    \label{fig:ablation}
  \end{center}
  \vskip -0.1in
\end{wrapfigure}

\textbf{Impact of the number of the attackers.}
Some previous works on untargeted model poisoning assume that there is a large fraction of attackers among participants~\cite{Fang2019LocalMP, ipm}. Therefore, we also investigate the number of attackers' impact on defense performance. In this section, we conduct experiments on non-iid partitioned MNIST with $q = 0.5$, and vary the number of attackers from 10\% to 40\% of the total participants. Results of this study are provided in Figure~\ref{fig:ablation}, showing the final global model accuracy of defense methods under various settings. We observe that the two defense baselines remain robust when the ratio of attackers is less than 30\%. As the ratio reaches 30\%, both Krum and Clipping Median collapse on the accuracy performance and fail to defend the model convergence, while \texttt{CosDefense} still provides robust accuracy performance. One notable point is that the \texttt{CosDefense} achieves the best accuracy compared to other baselines among all different ratios of attackers. The experiment results demonstrate that \texttt{CosDefense} is robust and reliable under both small-scale and large-scale attack 
scenarios.


\section{Conclusion}
We presented \texttt{CosDefense}, a cluster-based defense scheme that could filter the malicious updates out of the model aggregation on the server. Experiment results on real-world datasets demonstrate that \texttt{CosDefense} can provide robust performance under the state-of-the-art FL poisoning attack. We picture the limitations and potential future work in Appendix~\ref{app:future}.

\section{Acknowledgement}
This material is based upon work supported by Defense Advanced Research Projects Agency (DARPA) under Contract No. HR001120C0156, ARO award W911NF1810400, ONR Award No. N00014-16-1-2189. The views, opinions, and/or findings expressed are those of the author(s) and should not be interpreted
as representing the official views or policies of the Department of Defense or the U.S. Government.

\bibliography{iclr2023_conference}
\bibliographystyle{iclr2023_conference}

\clearpage
\appendix
\section{Preliminary}
\subsection{Federated Learning}
In this paper, we consider the federated learning setting that is similar to the vanilla FedAvg~\cite{fedavg}, in which the FL system is composed of a server with $\mathcal{K}$ clients, whose data is only locally kept without sharing. The clients cooperate in training a global model $\mathcal{W}_g$ with parameters $\theta$. We consider the following distributed optimization problem: $\min\limits_{\theta} f(\theta)$, where $f(\theta) := \sum_{i=1}^{\mathcal{K}} p_i F_i(\theta)$. The $F_i(\cdot)$ represents the local objective of client $i$ and $p_i$ denotes the aggregation weight of client $i$ satisfying $p_i \geq 0$ and $\sum_{i=1}^{\mathcal{C}} p_i=1 $. 

The federated learning algorithm runs as follows: in each communication round $t$, the server randomly selects a subset $S_t$ from available clients for client sampling and sends the global model parameters $\theta_{t}$ to the selected clients. Each selected client $i \in S_t$ initializes its local model parameters as $\theta_{t}$ and performs local model updates $\theta^i_{t+1} = \theta_{t} - \eta \frac{1}{B} \sum_{z \in b_i} \nabla_\theta \ell(\theta_{t}; z)$, where $\eta$ is learning rate, $b_i$ is a mini-batch of size $B$ randomly sampled from local data $\mathcal{D}_i$. After local update finishes, each client $i \in S_t$ send the model update $g^i_{t+1} := \theta^i_{t+1} - \theta_{t}$ to the server. Lastly, the server performs the aggregation function over the received model updates and obtains a new global model such that  $\theta_{t+1} \leftarrow \theta_t - G_t$, where $G_t \leftarrow Aggr(\{g^i_{t+1}\}_{i \in S_t}) $. The newly updated model parameters $\theta_{t+1}$ is then used to perform the next round of FL training. The FL training is ended when the round number reaches $\mathcal{T}$.

\subsection{Threat Model}

\begin{figure}[h]
	\centering
 \vspace{2mm}
	\includegraphics[width=0.8\linewidth]{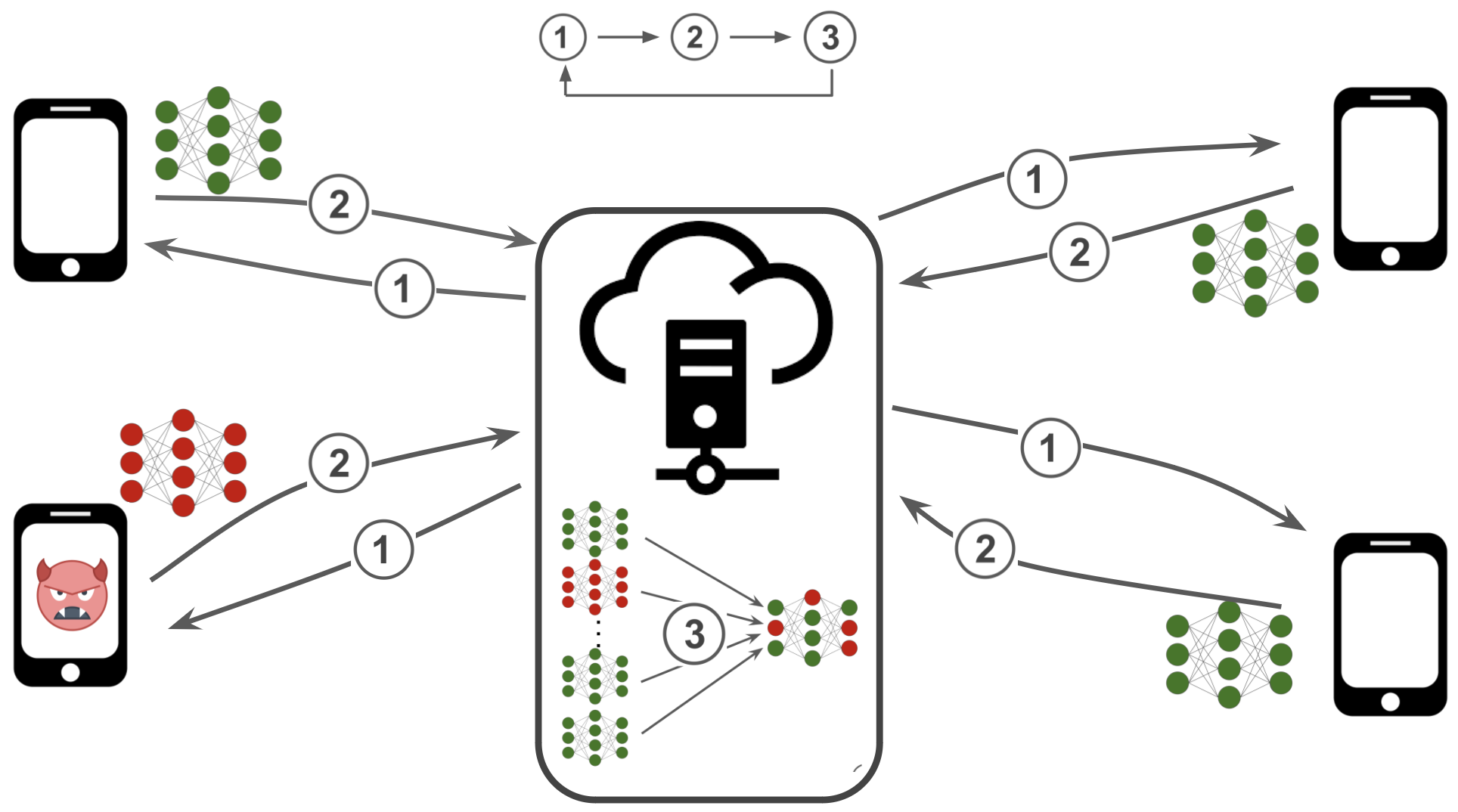}
    \caption{Threat model of the data poisoning attack in FL. After receiving the global model from the server, the malicious clients purposely upload contaminated model updates to poison the model.}
    \label{fig:setup}
\end{figure}

We assume that the attacker's goal is to diminish global model accuracy regardless of specific inputs or classes, in which malicious clients perform untargeted attacks. In this work, we assume that $p$ percentage of the total number of clients is malicious, where $0 \leq p \leq 1$. Let the malicious clients be denoted by $\mathcal{M}$ and the set of malicious clients as $\mathcal{A}$. All malicious clients share the same purpose and perform the same type of attack. They are either coordinated by an external attacker or by a leader client. During the FL training, the malicious attackers send crafted local updates $\left \{ \widetilde{g}^t_{k} \right \} _{k\in \mathcal{A} }$ to the server to purposely maximize the global optimization function $f(\theta)$. The crafted local updates are generated from their local data distributions, which differ from the benign clients' local distributions. The illustration of the threat model is shown in Figure~\ref{fig:setup}.

\section{Related Works}
\textbf{Poisoning Attacks in FL.} 
In federated learning, the attackers may control a fraction of the clients and manipulate the local data or training process to purposely poison the global model performance by injecting the wrong local model update. This kind of attack is called the model poisoning attack, in which the attacker contaminates the global model. 
Poisoning attacks are mainly categorized into targeted and untargeted attacks according to the attacker's aim. Targeted attacks aim to keep global model performance high, while the model considerably misbehaves on a specific set of inputs. This can be done either by injecting backdoor triggers to the inputs regardless of the original class or flipping the labels of only one class to a similar label (e.g., flipping 1 to 7 in MNIST or 'cat' to 'dog' in CIFAR10). 
On the other hand, the purpose of untargeted attacks is to decrease the global model's performance significantly. Label-flip \cite{sym-flip}, and  gradient modification (e.g., sending random noise to the server instead of real update) \cite{Krum} attacks are examples of this type. Moreover, \cite{Rl-attack} utilizes reinforcement learning to create an optimal untargeted attack, however, building the framework to generate online attack is costly in terms of both time and computation. Another state-of-the-art untargeted model poisoning attack is Inner Product Manipulation (IPM) proposed by \cite{ipm}. The aim of the IPM is to modify the local updates of the malicious clients such that the inner product between the true gradient and the aggregated updates is negative.


\textbf{Defenses for Poisoning Attacks in FL.} Several solutions have been proposed to mitigate the effect of the attacks, so that model performance is not negatively affected. Client-side defenses modify the local training algorithm and rely on benign clients \cite{FAT, FL-WBC}. For example, \cite{FL-WBC} finds the model parameters that the attack is hidden and perturbs those parameters during local training of benign clients. Some server-side defense methods decrease the aggregation weights of the malicious client updates \cite{ ByGARS, FLTrust,  RFFL}. While others perform clustering based on different techniques and aggregate only benign clients \cite{Krum, PCA}. For instance, Multi-Krum \cite{Krum} calculates the total Euclidean distance from the $n-f-2$ nearest neighbors for each client update, where $n$ is the number of clients and $f$ is the estimated malicious client number. Then, only the $n-f$ updates with the minimum distance is aggregated. Another work \cite{DP2} applies norm-clipping to the client updates before the aggregation, based on the observation that norm of malicious updates are large. Also, differential privacy methods (e.g., adding noise to updates) are shown to be mitigating model poisoning attacks \cite{DP2, DP1}. 

However, some current solutions have critical drawbacks in terms of real-world feasibility. For instance, algorithms proposed by \cite{ByGARS, FLTrust, Sageflow} rely on representative benign data on the server, which is hard to accomplish in actual FL deployment, where the server seldom stores the ground-true validation data. Moreover, \cite{RFFL} is incompatible with client selection, an inevitable necessity for cross-device FL where thousands of clients participate. Also, Krum and Multi-Krum \cite{Krum} require the knowledge of the number of malicious clients in the FL setting , which is not feasible. Lastly, methods proposed by \cite{RFFL, FoolsGold, ByGARS} maintains reputation scores for each client on the server side, which is a serious overhead since thousands of clients may participate in the FL system. Overall, there is a gap for a defense solution in the FL setting that does not require server data, is compatible with client selection, and is self-sufficient for detecting malicious clients.

\section{Algorithm} \label{algorithm}
\texttt{CosDefense} algorithm is presented in the Algorithm \ref{alg:FedDetect}. More specifically, after calculating the cosine similarity scores of each client, the server takes their absolute value before and performs min-max normalization, which is beneficial for separating the actual malicious and benign clients better. After these two adjustments, we define the threshold value as the mean of current scores and cluster clients accordingly. If the client score exceeds the mean value, it is labeled malicious. Otherwise, the server considers it benign and aggregates its weights for the next global model. 
\begin{algorithm}
   \caption{\texttt{CosDefense}}
   \label{alg:FedDetect}
\begin{algorithmic}
   \STATE {\bfseries Input:} Global model parameters $\theta_t$,  local updates $g^i_{t+1}$ of sampled clients $i \in S_t$, and aggregation method $Aggr(\cdot)$
   \STATE {\bfseries Initialize:} Benign client set $B=\{\}$ß
   \FOR{each client $i$ in $S_t$}
   \STATE $cs_i \leftarrow |cosine\ similarity(\theta_{t, L}, g_{t+1, L}^i)|$
   \ENDFOR
   \FOR{each client $i$ in $cs_i$}
        \STATE $cs_i \leftarrow (cs_i - min(cs) ) / (max(cs) - min(cs))$
   \ENDFOR
   \STATE threshold $\leftarrow mean(cs)$
   \FOR{each client $i$ in $S_t$}
        \IF{$cs_i <$ threshold}
        \STATE $B \leftarrow B \cup \{i\}$ 
        \ENDIF
   \ENDFOR
   \STATE $G_g \leftarrow Aggr(\{g_{t+1}^i\}_{i\in B})$
   \STATE $\theta_{t+1} \leftarrow \theta_t - G_t$
   \STATE {\textbf{Output:} $\theta_{t+1}$} 
\end{algorithmic}
\end{algorithm}

\section{Experimental Details} \label{app:exp_details}

\textbf{Datasets and partitioning.}
Dataset statistics are provided in Table \ref{tab:dataset}. We generate the non-iid partition for all datasets following previous FL works~\cite{Rl-attack, Fang2019LocalMP}. Specifically, suppose there are $C$ various classes in a dataset. We evenly break all the clients into C groups, in which each group is assigned $1/C$ of training samples as follows. A training point with label $c$ would be assigned to $c$-th group with probability $q \geq 1/C$ and to each of the rest groups with probability $(1-q)/(1-C)$. The attacking clients would be evenly distributed across $C$ groups. A higher $q$ value stands for a higher non-iid degree. In the experiments above, we set the default $q$ equals to 0.5 and partition the datasets into 100 clusters.

\begin{table}[h]
\centering
\scriptsize
\caption{
Dataset statistics.
}
\vspace{2mm}
\begin{tabular}{cccccc} \hline
\toprule
\textbf{Dataset} & \textbf{Train Clients} & \textbf{Train Examples} & \textbf{Test Examples}  & \textbf{Input Size}\\ 
MNIST & 100 & 60,000 & 10,000 & 1x28x28\\
Fashion-MNIST & 100 & 60,000 & 10,000 & 1x28x28 \\
CIFAR-10 & 100 & 50,000 & 10,000 & 3x32x32 \\
\bottomrule
\end{tabular}
\label{tab:dataset}
\end{table}

\textbf{Baseline implementations and parameter selections.}
We follow the same implementation details for Krum and Clipping-Median as in the previous work~\cite{Rl-attack}. 
For Krum, we set the number of the attacker parameter $f$ to its real value in each round for the best performance. For the Clipping-Median, we first apply norm-clipping \cite{DP2} then perform coordinate-wise median \cite{trimmed-mean}, since it is shown to be more powerful than the vanilla median \cite{Rl-attack}. Moreover, we provide the results for no defense case, where the server only performs FedAvg aggregation. 
For all experiments, the communication rounds fix to 1000, and the local training iteration number is 1 with batch size 128. The aggregation function is FedAvg \cite{fedavg}. The default local learning rate is 0.01 for all experiments. 

\textbf{Configurations.} 
All experiments are conducted by CPU/GPU simulation. The simulation experiments are conducted on a computing server with eight GPUs. The server is equipped with AMD EPYC 7502 32-Core Processor and 1024G memory. The GPU is NVIDIA RTX A4000.

\section{Experimental Results} \label{app:result}

The final global model accuracy of each method is provided in the table \ref{tab:acc}.

\begin{table}[h]
\centering
\scriptsize
\caption{
Global model accuracy comparison between No Defense, Krum, Clipping-Median, and \texttt{CosDefense}.
}
\vspace{2mm}
\begin{tabular}{ccccc} \hline
\toprule
 & No Defense & Krun & Clipping-Median & \texttt{CosDefense} \\ 
\midrule
MNIST & 9.8\% & 9.8\% & 19.29\% & \textbf{83.37\%}\\
\midrule
F-MNIST & 10\% & 10\% & 21.24\% & \textbf{67.63\%}\\
\midrule
CIFAR-10 & 10\% & 10\% & 20.89\% & \textbf{52.02\%}\\
\midrule

\end{tabular}
\label{tab:acc}
\end{table}

\begin{figure}[h]
	\centering
 \vspace{2mm}
	\includegraphics[width=0.45\linewidth]{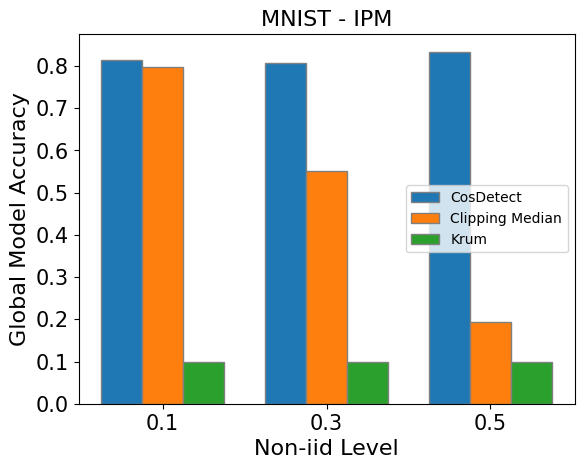}
    \caption{Accuracy performance on MNIST dataset under different non-iid partition levels.}
    \label{fig:ablation2}
\end{figure}

\textbf{Impact of the non-iid degree.}
In order to investigate the impact of data heterogeneity on \texttt{CosDefense}, we partition the MNIST dataset with $q$ equals 0.1, 0.3, and 0.5, respectively, and run all three defense strategies on top of the IPM attack for the evaluation. We set 30\% of the devices as malicious clients, the same as in the previous section.
As shown in Figure~\ref{fig:ablation2}, we observe that \texttt{CosDefense} is stable with the change of the non-iid degree, and outperforms the other two defense baselines on accuracy performance among all scenarios. 
The results indicate that the \texttt{CosDefense} is resistant to the impact of data heterogeneity compared to the other defense baselines.


\section{Limitations and Future works.} \label{app:future}
One of the directions for future work is the combination of \texttt{CosDefense} with the existing secure aggregation frameworks. The challenging point is that under the secure aggregation frameworks, each model update is locally encrypted and uninspectable by a server. Instead, only the sum of the model update would be revealed to the server when sufficient updates have been received, which impedes the implementation of \texttt{CosDefense} on top of it.

\end{document}